\documentstyle[aps,preprint]{revtex}

\newcommand{\be}{\begin{equation}}   \newcommand{\ee}{\end{equation}}
\newcommand{\ba}{\begin{array}}      \newcommand{\ea}{\end{array}}
\newcommand{\bea}{\begin{eqnarray}}  \newcommand{\eea}{\end{eqnarray}}
\begin{document}
\title{Dynamical Renormalization Group Study of a Conserved Surface Growth
 with Anti-Diffusive and Nonlinear Currents }
\author{Youngkyun Jung$^{\dagger}$,
In-mook Kim$^{\dagger*}$
and Yup Kim$^{\ddagger**}$}

\maketitle
\centerline{ $\dagger$   {\em Department   of Physics,   Korea University,   Seoul, 136-701, 
KOREA}}
\begin{center}
$\ddagger$ {\em Department of Physics and
Research Institute for Basic Sciences \\ Kyung-Hee University
Seoul 130-701, KOREA}
\end{center}

\date{~}
\medskip

\begin{abstract}
Based on dynamical renormalization group (RG) calculations to
the one-loop order, the surface growth described by a nonlinear
stochastic conserved growth equation, ${\partial h
\over \partial
t}= \pm \nu_2\nabla^2h
+\lambda\nabla\!\cdot\!(\nabla h)^3+\eta \; \; (\nu_2>0)\/$,
is studied analytically.
The universality class of
the growth described by the above equation with
$+\nu_2\/$ (diffusion) is shown to be the same as
that described by the Edwards-Wilkinson (EW) equation
(i.e. $+\nu_2\/$ and $\lambda=0\/$).
In contrast our RG recursion relations manifest that
the growth described by the above equation
with $-\nu_2\/$(anti-diffusion) is an unstable growth and
do not reproduce the recent results from a numerical
simulation by J. M. Kim [Phys. Rev. E {\bf 52}, 6267 (1995)].
\draft
\vspace{0.5 cm}

PACS No.:\ 05.40.+j; 05.70.Ln; 68.35.Fx;  61.50.Cj.

\vspace{0.5cm}
$^*$ E-mail : imkim@kuccnx.korea.ac.kr

$^{**}$E-mail : ykim@nms.kyunghee.ac.kr

\end{abstract}

\newpage
Recently, there have been considerable interests in the various surface
growth models \cite{dyn} to understand roughenings
in the growing surfaces theoretically.
Since the  surface structures of many growth process are self-affine,
most efforts have concentrated on the surface width $W$, which
is defined by the root mean square fluctuation of the surface height.
In a finite system of lateral size $L$,
the width $W$ starting from a flat substrate scales
as \cite{famvic}
\bea
W (t)& \sim & L^\alpha f({t/L^z}) \nonumber \\
    & \sim & t^\beta , \ \ \ \ \ \ \   t\ll L^z  \\
    & \sim & L^\alpha , \ \ \ \ \ \ \    t\gg L^z  \nonumber
\eea
where the scaling function $f(x)$ is $x^\beta$ for $x\ll 1$ and
constant for $ x\gg 1$.
 The exponents $\beta $ and $z$ are
connected by the relation  $z =\alpha/\beta$.
Among the growth models which satify Eq. (1),
the class of models known as ``conserved'' models
\cite{wolfvil,laisar,wilvz,vve,tambsar,yup,jmsarma},
which
conserve the total number of particles after being deposited,
has
been extensively studied as a possible description for the real
molecular beam epitaxy (MBE) growth \cite{he}.
In these conserved models
the height $h$ describing the local position of the surface
of growing materials is known to obey
\cite{wolfvil}
\be
\frac{\partial h({\bf x},t)}{\partial t}=
                    -\nabla \cdot {\bf J}({\bf x},t)+\eta({\bf x},t)~~,
                   \label{eq:conservation}
\ee
where $\rm {\bf J}$ is the surface current and
$\eta ({\bf x},t)$
is an uncorrelated random noise
\be
\langle \eta({\bf x},t) \eta({\bf x}',t') \rangle =
                   2D\delta^d ({\bf x-x'}) \delta(t-t')~~.
                        \label{eq:noise}
\ee

There have been several studies on the
surface growth described by the following current
\cite{jmsar,sarkot,jm,sarlkg,kshgh,ryuin}
\be
{\bf J}({\bf x},t)= -\nu_2 \nabla h + \nu_{4}\nabla^3 h
                      -\lambda(\nabla h)^{3}
                    \label{eq:current}
\ee
and thus by the corresponding continuum equation
\be
\frac{\partial h({\bf x},t)}{\partial t}=
        \nu_2 \nabla^2 h - \nu_4 \nabla^4 h
       +\lambda \nabla \!\cdot\! (\nabla h)^3 + \eta({\bf x}, t)~~.
                  \label{eq:continuum}
\ee

For $\nu_2 > 0$, the critical property of the surface growth described
by Eq. (\ref{eq:continuum})
is solely dependent on the first term
$\nu_2\nabla^2 h$ in the right hand side, which is
the most relevant term in a renormalization group (RG) sense and
it belongs to the Edwards and Wilkinson(EW) universality class \cite{ew}
with $\alpha =(2-d)/2$ and $z=2$
for any substrate dimension $d\/$ \cite{jmsar,sarkot}.
When $\nu_2 = \lambda =0\/$,  Eq. (\ref{eq:continuum}) becomes
the Mullins-Herring equation \cite{hermul},
which can be solved exactly
to give $\alpha = (4-d)/2\/$ and $z=4\/$ \cite{wolfvil}.
When $\nu_2=0,~\nu_4>0 \ $ and $\lambda > 0$,
the recent studies \cite{jmsar,sarkot}
have shown that
the most relevant nonlinear $\lambda$ term
generates an effective $\nu_2^{eff} \nabla^2 h $ term with
$\nu_2^{eff} > 0 $ and the
corresponding growth belongs to the EW
universality class, even though
both the scaling argument and the dimensional analysis \cite{laisar}
suggest the exponent values $\beta =(4-d)/2(4+d)\/$
and $\alpha = (4-d)/4\/$.

More recently the growth described by
Eq. (5) with $\nu_2 < 0 $,  $\nu_4 \ge 0 $ and $\lambda > 0$
has been studied numerically \cite{jm}. From this numerical
results it has been argued that
the critical property of such growth
belongs to the EW universality class even for $\nu_2 < 0 $, on the ground
that the nonlinear $\lambda\/$ term could
supress any negative $\nu_2\/$, so that $\nu_2\/$ could become effectively
positive, i.e. $\nu_2^{eff} > 0$ \cite{jm}. If this argument is right, then the
$\lambda$ term could make the unstable growth with -$|\nu_2| \nabla^2 h$ term
the stable EW growth with $\nu_2^{eff} > 0$.
However if one analyzes Eq. (\ref{eq:continuum}) from a point of view of
the scaling and dimensional
analyses, the most relevant term
is $\nu_2 \nabla^2 h $ regardless of the sign of $\nu_2$.
So it is somewhat strange that the less relevant $\lambda$ term renormalizes
the most relevant $\nu_2$ term so effectively as to make the change
of sign of $\nu_2$. It is thus our motivation to check
whether this strange renormalization is physically plausible or not
by use of dynamical renormalization
group calculations \cite{fns,mhkz} and other analytical methods.

To achieve this goal, we first discuss the dynamical renomalization
group study of Eq. (\ref{eq:continuum}) with
$\nu_2 \neq 0 $, $\nu_4=0$ and $\lambda>0$. In the large distance and
long-time hydrodynamic limits,
Eq.~(\ref{eq:continuum}) for $\nu_4=0\/$ in Fourier space
can be written as
\bea
h({\bf k},\omega) &=& G({\bf k},\omega) \eta({\bf k},\omega) \nonumber \\
&=& G_{0}({\bf k},\omega)\eta({\bf k},\omega)+
\lambda G_{0}({\bf k},\omega) \nonumber \\
& &\times \int \!\!\! \int \!\!\! \int \!\!\! \int
\frac{d\Omega_1 d^{d}{{\bf q}_1}d\Omega_2 d^{d}{{\bf q}_2}}
{(2\pi)^{2d+2}}\nonumber\\
& &\times ({\bf q}_1 \cdot {\bf q}_2 )[{\bf k}
\cdot ({\bf k}\!-\!{\bf q}_1 \!-\!{\bf q}_2 )] h({\bf q}_1 , \Omega_1)\nonumber\\
& &\times h({\bf q_2}, \Omega_2)
h({\bf k}\!-\!{\bf q}_1\!-\!{\bf q}_2 , \omega \!-\! \Omega_1 \!-\! \Omega_2)
                         \label{eq:Fourier}
\eea
where $G_{0}({\bf k},\omega)\/$ is the bare propagator defined
by the expression
\be
G_{0}({\bf k}, \omega) = \frac{1}{\nu_2 k^2 -i\omega}~~.
                       \label{eq:propagator}
\ee
Using $\langle\eta({\bf k},\omega) \eta({\bf k}',\omega') \rangle
=2D\delta^d ({\bf k}+{\bf k}')
\delta(\omega+\omega')\/$
and performing the internal frequency integrals of Eq. (\ref{eq:Fourier}) to
the one-loop order,
one obtains
\be
G({\bf k},\omega)=G_0({\bf k},\omega)-\frac{\lambda D}{\nu_2}
               K_d \frac{d+2}{d}k^2 G^2_0({\bf k},\omega)\Sigma(q),
\ee
in the limits $\omega \rightarrow 0\/$
and $k \rightarrow 0\/$. Here $\Sigma(q)=\int^{\Lambda}dq q^{d-1}\/$,
$\Lambda\/$ is the momentum cutoff $(\Lambda \equiv 1)\/$,
and $K_{d}=S_{d}/(2 \pi)^{d}\/$
with $S_{d}=2\pi^{d/2}/ \Gamma (d/2)\/$.
Since $\Sigma(q)$ has no infrared divergence
for any dimension, we can expect that the nonlinear
$\lambda$ term is irrelevant in RG
sense and
that the $\lambda$ will not renormalize negative $\nu_2$ term
to be positive.

For the confirmation's sake, we want to calculate the dynamical
RG recursion relations for $\nu_2$ and
$\lambda$.
Integrating out the fast modes in the momentum shell
$k^{>} \in [\Lambda e^{-\ell}, \Lambda]\/$ and restoring the slow
modes $k^{<} \in [0, \Lambda e^{-\ell}]\/$ with
$e^{-\ell}=1-\delta \ell+\cdot \!\! \cdot \!\! \cdot\/$,
we find
an effective surface tension $\nu^{<}_{2}\/$ for the long wavelength modes,
\be
\nu^{<}_{2} = \nu_2 \left [ 1+\delta \ell K_{d}\frac{\lambda D}
            {\nu^{2}_2} \frac{d+2}{d} \right ]
                           \label{eq:nu_2}~~.
\ee
In an analogous way we get
\be
\lambda^{<} = \lambda \left [ 1-\delta \ell K_{d}a(d)
                        \frac{\lambda D}{\nu^{2}_2} \right ]
                     \label{eq:lambda}~~,
\ee
where $a(d)=9\/$ for $d=1$,
and $a(d)=\frac{9}{2}\/$ for $d=2$.
Upon requiring that the equation stays invariant under
the scale($e^{\ell}$) transformations ${\bf k}\rightarrow e^{-\ell}{\bf k}$,
$t\rightarrow e^{z\ell}t$ and $h\rightarrow e^{\alpha\ell}$,
the parameters
should transform as follows:
$\nu_2 \rightarrow  b^{z-2}\nu_2$,
$D \rightarrow b^{z-2\alpha -d}D$ and
$\lambda \rightarrow  b^{z+2\alpha -4}\lambda$.
Combining these scale transformations
and Eqs. (\ref{eq:nu_2}) and (\ref{eq:lambda}),
we get the following RG recursion relations:
\bea
\frac{d \nu_2}{d\ell} &=& \nu_2 \left[ z-2+gK_{d}
                      \frac{d+2}{d} \right] \label{eq:nu_2d}\\
\frac{dD}{d\ell} &=& D \left[ z-2\alpha-d \right] \label{eq:Dd}\\
\frac{d \lambda}{d\ell} &=& \lambda \left[ z+2\alpha-4-ga(d)K_{d}
                \right]~~, \label{eq:lamd}
\eea
where the coupling constant $g$ is defined as
\be
g\equiv\frac{\lambda D}{\nu_2^2}~~. \label{coco}
\ee
The RG recursion relation of the effective coupling constant $g$ is thus
\be
\frac{dg}{d\ell}=g\left[-d-gA(d) K_d \right] ~~ ,
                 \label{eq:couplingdiff}
\ee
where $A(1)=15\/$ and $A(2)=\frac{17}{2}\/$
for $d=1$ and $2\/$ and we have also confirmed $A(d) > 0$ for any $d$.
As one can expect from Fig. 1,
the RG flow of $g$ has only one stable fixed point which
is a Gaussian fixed point, i.e. $g^*=0$ and does not
have any nontrivial fixed points.
The RG flow of $g$ for $\nu_2 <0$ is the same as
those for $\nu_2 > 0 $. If the results from the numerical
calculations in Ref. \cite{jm} would be explained by the
dynamical RG, then the RG flow of $g$ for $\nu_2 < 0$
should be different from those for $\nu_2 > 0$.
As shown in Fig. 2, the RG flows starting at
$(\nu_2 > 0, \lambda > 0)$ will eventually arrive
at $(\nu_2^{eff} >0, \lambda \rightarrow 0)$.
This means that the growth described by Eq. (5)
with $\nu_4=0$, $\nu_2 > 0$ and $\lambda > 0$
in the large distance scale
belongs to the EW universality class as we have expected.
The RG flows starting at
$(\nu_2 < 0, \lambda > 0)$, on the while, will eventually arrive
at $(\nu_2^{eff} < 0, \lambda \rightarrow 0)$.
This means that the growth described by Eq. (5) with
$\nu_4 = 0$, $\nu_2 < 0$ and $\lambda >0$
in the large distance scale
has nearly the same critical
property as the unstable linear growth described
by
\bea
\frac{\partial h({\bf x},t)}{\partial t}=
       -|\nu_2| \nabla^2 h + \eta
                  \label{eq:us}~~.
\eea
Based on our dynamical RG calculations
we conclude that the growth by Eq. (5) with $\nu_2 <0$,
$\lambda > 0$ and $\nu_4=0$
should be an unstable growth and we cannot reproduce
the recent results by a numerical simulation \cite{jm}.

The RG recursion relations (11), (12) and (13) are
based on the one-loop order 
perturbative calculations and 
are intrinsically exact only for $\lambda \ll 1$ or $g=\frac{\lambda D}{\nu_2^2} \ll 1$.
The main result of the dynamical RG calculations physically indicates that
the growth described by Eq. (5) with $\nu_2 <0$, $\nu_4=0$ and the small $\lambda(> 0)$ 
does not belong to the same universality class as that of the EW growth.
To understand physically the unstable growth by Eq. (5)
with $\nu_2 <0$, $\lambda > 0$ and $\nu_4=0$
for the finite $\lambda$ including the case
for $\lambda \ll 1$, let's discuss the growth  
from an Hamiltonian-based argument or from an equlilibrium physics.
If one believe that the dynamical equation (\ref{eq:continuum}) with $\nu_4=0$
can be derived by an
equilibrium Hamiltonian ${\cal H}$ via the Langevin equation ${\partial h
\over \partial
t}= -\frac{\delta {\cal H}}{\delta h}+\eta $ and the Langevin
equation will reach a steady state or an equilibrium
where the states are controlled
by the Boltzman distribution
$P({\cal H}) \propto exp (-\beta {\cal H})$,
${\cal H}$ should be
\bea
{\cal H} = \int d^d x ~[ \frac{\nu_2}{2} (\nabla h)^2 ~+~ \frac{\lambda}{4}
(\nabla h)^4 ] \label{eq:Ha}~~.
\eea
If one puts $\nabla h$ to be equal to
a field $\phi(x)$ (i.e.,
$\nabla h \equiv \phi(x)$ ) and use a mean-field theoretic argument
for Hamiltonian (\ref{eq:Ha}), the corresponding Landau-Ginzburg function
${\cal L}$ becomes
\be
{\cal L} = \frac{\nu_2}{2} \phi^2 ~+~\frac {\lambda}{4}
\phi^4   \label{eq:La}~~.
\ee
When $\nu_2 > 0$ and $\lambda > 0$, the mean field theory with
$\frac{\partial {\cal L}}{\partial \phi} = 0$ gives
$\langle \phi \rangle=\langle \nabla h \rangle=0$ as in a disordered phase
of the ordinary magnetic phase transitions \cite{fli}.
The fluctuations around the mean-field
$\langle \phi \rangle=\langle \nabla h \rangle=0$
should be a Gaussian-type and it seems quite plausible to
believe that the surface roughenings with
$\nu_2 > 0$ and $\lambda > 0$ belong to the EW universality class.
However when $\nu_2 < 0 $ and $\lambda > 0$, the mean field thory predicts
\be
\langle\phi^2\rangle=\langle(\nabla h)^2\rangle_{eq}= - \frac{\nu_2}{\lambda} > 0
\label{eq:or}
\ee
as in an ordered phase of the ordinary magnetic phase transitions.
In the surface growth phenomena, Eq. (\ref{eq:or})
means that the local slope satisfies
$\langle|\nabla h| \rangle_{eq} \simeq \sqrt
{|\nu_2| / \lambda}$ in an equilibrium state, i.e. in a saturation state.
Dynamically, in the growing process when $\langle|\nabla h|\rangle
< \sqrt{|\nu_2| / \lambda}$,
the anti-diffusion term $-\nu_2 \nabla h$
of the current (see Eq. (\ref{eq:current})) dominates.
Dynamical RG recursion relations (11), (12) and (13) which are
intrinsically exact for $\lambda \ll 1$ or $g=\frac{\lambda D}{\nu_2^2} \ll 1$
correspond to this growing stage of the local slopes. 
For $g \ll 1$ where RG calculations are correct, it takes very long time for the local slopes
to reach the equilibrium value $\sqrt {|\nu_2| / \lambda}$ and this dynamical process
should be described by Eq. (16), because $\langle|\nabla h| \rangle_{eq} \simeq \sqrt
{|\nu_2| / \lambda} \gg 1$. We have confirmed this kind of growth
by using several simulations for the growthes with different $g$'s \cite{fra}.
In contrast for finite $g$'s the time interval in which the anti-diffusion dominates 
become rather finite and sooner or later the growth reaches the saturation state 
or the equilibrium state where
the effect of the anti-diffusion term is balanced to give $\langle|\nabla h|\rangle
\sim \sqrt{|\nu_2| / \lambda}$. After that local slopes flucuates around the value in Eq. (19). 
We have also confirmed that this kind of the unstable growth
occurs when $g$ is finite by simulations \cite{fra}.
It is thus not physically sound that the growth with $\nu_2 < 0$ and $\lambda > 0$
belongs to the same universality class as that of the EW-like growth
with $\nu_2^{eff} \nabla^2 h$ with $\nu_2^{eff} > 0$ as in Ref \cite{jm}.
Instead we do expect an unstable growth. 

We now want to discuss on the point why
the numerical simulations in Ref.
\cite{jm} couldn't see such an unstable
growth and did see only the stable
growth which belongs to the EW universality class.
To do a numerical simulation in Ref. \cite{jm},
the local currents in Eq. (\ref{eq:current})
in $d=1$ has been set to be equal to
\bea
\hat j(k,i)&=&\nu_2 [h(k)-h(i)] - \nu_4 [h(k+1)+h(k-1) \nonumber \\
          & &-2h(k)-h(i+1)-h(i-1)+2h(i)] \nonumber \\
          & &+\lambda [h(k)-h(i)]^3~~,   \label{eq:sim1}
\eea
where
$k$ is either $i+1$ and $i-1$ and a particle is added
on site $k$ if $\hat j (k)$ is negative.
If both $\hat j(i+1,i)$
and $\hat j(i-1,i)$ are positive,
a particle is added on   site $i$. In case both    $\hat j(i+1,i)$ and
$\hat j(i-1,i)$   are
negative, a particle is added on either site randomly.
When $\nu_4=0$, Eq. (\ref{eq:sim1})
can be written as
\bea
\hat j(k,i)=\nu_2 [h(k)-h(i)]\{1+\frac{\lambda}{\nu_2} [h(k)-h(i)]^2\}~~.
\label{eq:sim2}
\eea
In the initial growing process
with $\nu_2 < 0$
when $\langle[h(k)-h(i)]^2\rangle$ is quite small
and \{ \} term in Eq. (\ref{eq:sim2}) is positive,
the anti-diffusion should
dominate the growth. As the growth is developed,
$\langle[h(k)-h(i)]^2\rangle$ increases to
$-\nu_2/\lambda$ rapidly and after
that \{ \} term in Eq. (\ref{eq:sim2})
fluctuates around zero.
We have confirmed that at this stage the local slope
$\langle|h(k)-h(i)|\rangle$ is nearly equal
to $\sqrt{-\nu_2/\lambda}$
\cite{fra}.
In contrast the numerical simulations
in Ref. \cite{jm} have been done only for the cases
$-\nu_2/\lambda =1,3$ and
the corresponding saturated local slopes
$|h(k)-h(i)|$ are only 1 and $\sqrt{3}$. Since $h(k)$
must be integer numbers 0,1,2,... in a simulation,
$|h(k)-h(i)| = 0,1$ or 2 could be
easily generated even by the noise effect only,
so the \{ \} term in Eq. (\ref{eq:sim2}) being negative
for $-\nu_2 /\lambda = 1,3$.
This fact explains why the numerical simulation
in Ref. \cite{jm} had seen only the EW-type
growthes instead of the unstable growth.
But for the considerably large $-\nu_2/\lambda$,
the anti-diffusion term dominates from the beginning
and this effect stays longer and longer until
being cancelled out by the nonlinear $\lambda$ term.
Hence hardly realizes the EW-like behavior.

\begin{center}
{\bf Acknowledgement}
\end{center}
The authors wish to acknowledge the critical and fruitful discussions
with Dr. Jin Min Kim. This work is supported in part by
the KOSEF (951-0206-003-2), the Ministry of Education (BSRI-96-2409),
and by the KOSEF through the SRC program of SNU-CTP.

\begin{figure}
\caption
{The RG flows of the coupling constant $g$.
  There is only one fixed point $g^{*}=0\/$ which is attractive
  fixed point. The RG flows of $g$ for $\nu_2<0$ are the same as
  those for $\nu_2>0$. (See Eq. (15))}
\label{Fig. 1}
\end{figure}

\begin{figure}
\caption
{Schematic RG flows in $(\lambda, \nu_2)\/$-plane.
for $d=1$ and $2\/$, respectively.
The RG flows starting at
$(\nu_2 > 0, \lambda > 0)$ will eventually arrive
at $(\nu_2^{eff} >0, \lambda \rightarrow 0)$.
In contrast the RG flows starting at
$(\nu_2 < 0, \lambda > 0)$ will eventually arrive
at $(\nu_2^{eff} < 0, \lambda \rightarrow 0)$.}
\label{Fig. 2}
\end{figure}

\end{document}